\documentclass[11pt,preprint,flushrt]{aastex}
\pdfoutput=1

\def\eqq#1{Equation~(\ref{#1})}
\newcommand\etal{{\it et al.\/}}
\newcommand\eg{{\it e.g.\ }}
\newcommand\ie{{\it i.e.\ }}

\input cyracc.def

\font\twlcyr=wncyr10 at 12pt

\def\cyr{\twlcyr\cyracc}

\def\shah{{\mbox{\cyr Sh}}}
\def\sinc{{\mbox{\rm sinc}}}
\def\boxcar{{\mbox{$\Pi$}}}
\begin{document}

\slugcomment{24 Feb 2014, Accepted to PASP}

\title{Resampling images in Fourier domain}
\author{Gary M. Bernstein\altaffilmark{1}, Daniel Gruen\altaffilmark{2,3}}
\altaffiltext{1}{Dept. of Physics and Astronomy, University of Pennsylvania,
Philadelphia, PA 19104}
\altaffiltext{2}{University Observatory Munich, Scheinerstrasse 1, 81679 Munich, Germany}
\altaffiltext{3}{Max Planck Institute for Extraterrestrial Physics, Giessenbachstrasse, 85748 Garching, Germany}

\email{garyb@physics.upenn.edu}

\begin{abstract}
When simulating sky images, one often takes a galaxy image $F(x)$
defined by a set of pixelized samples and an interpolation kernel, and then
wants to produce a new sampled image representing this galaxy as it
would appear with a different point-spread function, a rotation, shearing, or
magnification, and/or a different pixel scale.  These operations are
sometimes only possible, or most efficiently executed, as resamplings of the
Fourier transform $\tilde F(u)$ of the image onto a $u$-space grid that differs
from the one produced by a discrete Fourier transform (DFT) of the samples.
In some applications
it is essential that the resampled image be accurate to better than 1
part in $10^3$, so in this paper we first use standard Fourier
techniques to show that Fourier-domain interpolation with a wrapped
sinc function yields the exact value of $\tilde F(u)$ in terms
of the input samples and kernel.  This operation scales with
image dimension as $N^4$ and can be prohibitively slow, so we next
investigate the errors accrued from approximating the sinc function
with a compact kernel.
We show that these approximations produce a
multiplicative error plus a pair of ghost images (in each dimension)
in the simulated image. Standard Lanczos or cubic
interpolators, when applied in Fourier domain, produce unacceptable
artifacts.  We find that errors $<1$ part in $10^3$ can be obtained by
(1) 4-fold zero-padding of the original image before executing the
$x\rightarrow u$ DFT, followed by (2) resampling to the desired $u$
grid using a 6-point, piecewise-quintic interpolant that we design
expressly to minimize the ghosts, then (3) executing the DFT back to $x$ domain.
\end{abstract}


\section{Introduction}

In real images, one obtains a finite, pixelized (\ie\ sampled)
rendition of objects, for example the point spread function (PSF) from
stellar images, or galaxy images.  Many forms of subsequent analysis
require continuum representations of the objects or their Fourier
transforms, for example to resample the objects onto a new grid, or to
predict the appearance of the object after rotation, distortion, or
convolution with a new PSF.   Our principal concern is the validation of weak gravitational
lensing (WL) measurement methods: the intrinsic appearance of a galaxy
is slightly magnified or sheared by the gravitational deflection of
intervening (dark) matter, then the image is convolved with the PSF of
the atmosphere and optics.  WL software must estimate the lensing
effects given a sampled, noisy version of a galaxy ensemble, and must
do so to better than part-per-thousand accuracy to recover all of the
information available information about dark matter and dark energy
\citep{HTBJ,AmaraRefregier}.   
Simulated sky images used to test these methods must therefore have
fidelity at least this good in rendering the sheared, convolved
versions of realistic galaxies \citep[e.g.][]{great10,great3}.  

The PSF of a sky exposure is usually
estimated empirically, often by fitting stellar images with a model
consisting of an $N\times N$ grid of pixel values $a_{ij}$ and a specified interpolation
function $K_x$, such that the continuum representation is
\begin{equation}
F(x,y) \equiv \sum_{i,j=-N/2}^{N/2-1} a_{ij} K_x(x-i, y-j),
\label{sampled}
\end{equation}
\eg\ as done by the {\sc PSFEx} software \citep{psfex}.  To
recover the intrinsic shape of a galaxy that has been observed through
this PSF, the observed shape must be
corrected for the PSF, an operation that is most straightforwardly
done in the Fourier domain \citep{FDNT}, and requires part-per-thousand accuracy in
the PSF representation \citep{HTBJ, AmaraRefregier}.  One question,
therefore, is how to compute the Fourier transform
\begin{equation}
\tilde F(u,v) = \int dx\,dy\, F(x,y) e^{-2 \pi i(ux+vy)}.
\label{ft2d}
\end{equation}
of a PSF defined by
interpolation on a grid of samples.  A simple discrete Fourier
transform (DFT) is insufficient, as it represents a periodic PSF, and
does not include the effects of the interpolation function.

A second question arises when simulating the effect of weak
gravitational lensing on real galaxies.  This requires taking
pixelized images of the real galaxies, then calculating their appearance
after application of a lensing shear, convolution with a new PSF, and
sampling on a new pixel grid \citep[e.g.][]{shera}.  Rotation,
distortion, or an incommensurate re-pixelization require interpolation
in either real or Fourier space, and the application of convolution
favors a Fourier-domain solution.  If we need to deconvolve the
original image for its PSF, a Fourier-domain solution is strongly
favored.  Hence we ask: what schemes for
interpolation of the Fourier representation of the galaxy $\tilde
F(u,v)$  (between DFT
samples) are needed to produce a simulated sheared, resampled galaxy
at part-per-thousand level?  In answering the first question we will
find an exact method, but it requires sinc interpolation
that can take $O(N^4)$ operations.  We will search for approximate
solutions that instead require $O(K^2N^2)$ operations for some small
kernel size $K$, but still attain the desired precision in rendering.

We are motivated by the WL science, but there are of course many
applications for an accurate method of interpolating
images in Fourier domain from a discrete set of points to arbitrary
$u$ values.  

Our conventions for Fourier transforms and associated functions are in
the Appendix. 

\section{Fourier transform of an interpolated sampled image}
\label{ftinterp}
We assume that an object is correctly modeled as an interpolation of a
square, finite grid of values, so that its surface brightness $F(x,y)$
is defined by the finite set of values $a_{ij}$ for $-N/2\le i,j <
N/2$ as per \eqq{sampled}.  We assume $N$ is even; an odd-valued $N$
can be padded with another row and column of zeros.  
We will assume that the interpolation kernel has $K_x(0,0)=1$ and $K_x(m,n)=0$ at integer $m,n$ other than the origin---this is the condition that the interpolation agree with the input samples at the sampled locations.
We will assume that the kernel is even, $K_x(x,y)=K_x(-x,-y).$
$F$ is zero beyond a bounded region if $K_x$ is.   We wish to know the
Fourier transform (\ref{ft2d}).

For notational simplicity we solve the one-dimensional case, which extends easily to two or more dimensions.
\begin{eqnarray}
F(x) & \equiv & \sum_j a_j K_x(x-j) \\
 & = & (f \ast K_x)(x), \\
f(x) & \equiv & \sum_{j=-N/2}^{N/2-1} a_j \delta(x-j).
\end{eqnarray}
The $\ast$ indicates convolution, and $f(x)$ is the original sampled
function.  From the convolution theorem, we have $\tilde F(u) = \tilde
f(u) \tilde K_x(u)$.  If we recognize the discrete Fourier transform (DFT) of the $a_j$ as
\begin{equation}
\tilde a_k = \sum_{j=-N/2}^{N/2-1} a_j e^{-2\pi i jk/N}, \quad a_j = \frac{1}{N}
\sum_{k=-N/2}^{N/2-1} \tilde a_k e^{2\pi i jk/N}
\end{equation}
we can write
\begin{eqnarray}
\tilde f(u) & = & \frac{1}{N}\sum_{j=-N/2}^{N/2-1} \sum_{k=-N/2}^{N/2-1}\tilde
  a_k e^{2\pi i jk/N} e^{-2\pi i j u} \\
 & = & \frac{1}{N}\sum_{k=-N/2}^{N/2-1} \tilde a_k\sum_{j=-N/2}^{N/2-1} e^{2\pi
   i j v}, \quad v \equiv k/N-u \\
 & = & \sum_{k=-N/2}^{N/2-1} \tilde a_k K_u(u - k/N), \\
K_u(v) & \equiv & e^{\pi i v} \frac{ \sin
   \pi N v} { N\sin \pi v} = e^{\pi i v } \frac{
   \sinc\ Nv} { \sinc\ v}.
\end{eqnarray}
$\tilde f(u)$ is equal to the convolution of the
DFT, which exists at the points $u_k = k/N$, with a $u$-domain kernel
$K_u$.  We need to avoid confusing the two interpolation kernels now
involved in the problem: an $x$-domain interpolant $K_x$ that is
chosen {\it a priori} as the definition of our function $F(x)$; and
the $u$-domain interpolant $K_u$ that we derive as necessary to obtain
the exact value of $\tilde F(u)$.

The
factor $e^{\pi i v}$ in $K_u$ results  
  from our convention of the samples starting at $x=-N/2$ and being
  asymmetrically placed about the origin.  The exact expression
  for $\tilde F(u)$ is 
\begin{equation}
\tilde F(u) = \tilde K_x(u) \sum_{k=-N/2}^{N/2-1} \tilde a_k e^{\pi i
  \nu } \frac{ \sinc\ N \nu} { \sinc\ \nu}.
\label{exact}
\end{equation}
Note that the ratio of sines (or sincs) in $K_u$ is equal
to the result of wrapping the interpolant $\sinc(Nv)$ at period 1. 
$\tilde F(u)$ extends to infinity unless the $x$-domain kernel has a
compact transform $\tilde K_x(u)$.  In practice one will need to
truncate at some $u_{\rm max}$, in effect defining a band-limited kernel.

We now provide a recipe for the typical application in which one has
an image of a galaxy with some input PSF, sampled at unity pixel scale
to an $N\times N$ image.  One wants to output an image of exactly
the same
galaxy after deconvolving the input PSF, applying some affine
transformation, convolving with an output PSF, and resampling onto a
pixel scale $\Delta$.  Because the output DFT has to have some
finite dimension $M\times M$, we necessarily will be rendering an
image that has been folded with period $M\Delta$, hence it is
necessary to choose $M$ large enough that the folded flux is small
enough to ignore.  An exact answer is available via discrete Fourier
methods only when the output image is zero outside a bounded region.

The steps are:
\begin{enumerate}
\item Obtain the $\tilde a_{ij}$ from the input $a_{ij}$ via an $N$-point
  DFT to give $u$-space values on a grid of pitch $1/N$.  The
  operation count is $O(N^2 \log N)$ using FFT methods. 
  \label{dftstep}
\item Select appropriate $M$ and construct a grid of the output
  sampled frequencies $u^\prime_{mn}$ that will have pitch
  $1/M\Delta$. The real-space affine transformation will be equivalent to
  a linear transformation and phase change in Fourier domain, so each
  $u^\prime_{mn}$ will have a corresponding input $u_{mn}$.  
\item Assign to each $u^\prime_{mn}$ the value obtained from the exact
  interpolation to $u=u_{mn}$ of the input image defined by \eqq{exact}.  This will
  require $O(M^2 N^2)$ operations for the $K_u$ summation, plus $O(M^2)$
  operations to multiply by $\tilde K_x(u_{mn})$.  (One must be sure
  here to include in the summation any aliases of $u^\prime_{mn}$ that
  map back to frequencies $<u_{\rm max}$.)
  \label{convolvestep}
\item Divide the DFT by the transform of the input PSF at $u_{mn}$ to
  effect the deconvolution.  The operation count is $O(M^2)$.
\item Multiply by the transform of the output PSF evaluated at
  $u^\prime_{mn}$.  This is $O(M^2)$.
\item Execute the DFT back to the desired real-space grid, with $O(M^2
  \log M)$ operations.
\end{enumerate}
The limiting step is the $u$-space convolution (\ref{convolvestep}),
which will generally take $O(N^4)$ operations, although in some
circumstances the 2d convolution can be factored to yield $O(N^3)$,
still the slowest step.
We will examine below the
consequences of approximating this exact interpolation with a more compact kernel. 

\section{Interpolator accuracy}
Consider an interpolation kernel $K(t)$, with Fourier transform $\tilde K(\nu)$, that is used reconstruct the value of a sine wave $\exp(2\pi i \nu_0 t$) from samples at integral $t$ values to some non-integral $t$.  
The sinc function has the unique property that it interpolates without
error for any frequency $-1/2<\nu_0<1/2$.  We will assume more
generally that the interpolation kernel is symmetric, $K(-t)=K(t)$, such that $\tilde K(\nu)$ is real and symmetric, and is known to be exact at the interpolation nodes, \ie\ for integral arguments $j$,
\begin{equation}
\label{exactinterp}
K(j) = \left\{\begin{array}{cc}
1 & j=0 \\
0 & j\ne 0
\end{array}
\right .
\end{equation}
The interpolated reconstruction of the sampled sine wave is 
\begin{eqnarray}
R(\nu_0, t) & \equiv & \sum_{j=-\infty}^\infty K(j-t) e^{2\pi i \nu_0 j} \\
 & = & \left[ e^{2\pi i \nu_0 t} \shah(t) \right] \ast K(t)  \\
\Rightarrow \tilde R(\nu_0, \nu) & = &  \left[ \delta(\nu-\nu_0) \ast \shah(\nu)\right] \tilde K(\nu)\\
 & = & \sum_{j=-\infty}^\infty \tilde K(\nu) \delta(\nu-\nu_0-j) \\
\Rightarrow R(\nu_0, t) & = & \sum_{j=-\infty}^\infty \tilde K(\nu_0+j) e^{2\pi i (\nu_0+j) t} \\
 & = & e^{2\pi i \nu_0 t} \sum_{j=-\infty}^\infty \tilde K(\nu_0+j) e^{2\pi i j t}.
\end{eqnarray}
The prefactor is the correctly interpolated sine wave, so let us define an error function $E(\nu_0,t)\equiv  e^{-2\pi i \nu_0 t} R(\nu_0,t)-1$. 
Then this error function is
\begin{eqnarray}
\label{eu}
E(\nu_0, t) & = & \sum_{j=-\infty}^\infty \tilde K(\nu_0+j) e^{2\pi  i j t} - 1 \\
 & = & \sum_{j=-\infty}^\infty \tilde K(\nu_0+j) \left(e^{2\pi i j t}-1\right) 
\label{efunc}
\end{eqnarray}
In the second line we have made use of the fact that \eqq{exactinterp}
requires $\sum_j \tilde K(j+\nu_0)=1$.
The sinc filter has $\tilde K(\nu)=\boxcar(\nu)$ which vanishes for
$|\nu|>\frac{1}{2}$.  The $E$ function hence vanishes for
$|\nu_0|<\frac{1}{2}$, as expected.

\subsection{Background conservation}
\label{flat}
Many astronomical images have signals atop a large constant $(\nu=0)$
background.  It is hence important that the $x$-domain interpolant
$K_x$ have an error function satisfying $E(0,t)=0$, otherwise a
resampled image will have background fluctuations as $t$ varies.  Note
also that the same criterion dictates whether object flux will be
conserved when the interpolant is used to shift the samples by a
constant fraction $t$ of a pixel. 
Putting $\nu_0=0$ in \eqq{efunc}, the fractional background fluctuations will be
\begin{eqnarray}
E(0,t) & = & 2\sum_{j=1}^{\infty} \tilde K(j) \left(\cos 2\pi j t - 1\right) \\
 & \approx & 2 \tilde K(1) \left(\cos 2\pi t - 1\right).
\end{eqnarray}
From the first line we can see that any interpolant having $\tilde
K(j)=0$ for $j\ne0$ will conserve background level.  The
nearest-neighbor and linear filters (see Appendix for definitions)
satisfy this exactly since $\tilde K(\nu) = \sinc(\nu)$ and
$\sinc^2(\nu)$, respectively, and the polynomial interpolants are
designed to meet this criterion.  The Lanczos interpolants do not, however, meet this criterion.  In the second line 
we have assumed that we are using interpolants, like Lanczos, that attempt to approximate the band-limiting properties of the sinc filter, and will hence have $|\tilde K(1)|\ll 1$ and $|\tilde K(j)|\ll |\tilde K(1)|$ for $|j|\ge2$.  In this case we can see that
\begin{itemize}
\item The interpolated background error will oscillate as $\cos 2\pi t_0-1$, \ie\ will be worst for interpolation to the $t_0=0.5$ midpoint between pixels, and
\item the maximum fractional background error will be $-4\tilde K(1)$.
\end{itemize}
In practice, the Lanczos interpolant or any other can be normalized to
conserve a constant background via
\begin{eqnarray}
K(t) & \rightarrow & K(t) / (1+E(0,t)) \\
\label{conservex}
 & \approx & K(t) \left[ 1 - 2\tilde K(1)  \left(\cos 2j\pi t - 1\right) \right] \\
\Rightarrow \tilde K(\nu) & \rightarrow &
\left[1+2\tilde K(1)\right] \tilde K(\nu) - \tilde K(1) \left[ \tilde K(\nu+1) + \tilde K(\nu-1)\right].
\label{conserveu}
\end{eqnarray}
The effect of enforcing background conservation on the interpolant is hence to degrade slightly the band-limiting characteristic of the filter, adding $O(\tilde K(1))$ ``wings'' to the frequency responses that extend to $|\nu|=1.5$.

\subsection{Interpolation in Fourier domain}
In \S\ref{ftinterp} we inferred that a wrapped sinc interpolation on
the grid of DFT Fourier coefficients at $u_j=j/N$ is the correct way to
calculate the transform $\tilde F(u)$ at values of $u\ne u_j$.  Here
we examine the errors from approximating the exact $K_u$ with a smaller
interpolation kernel.  Let us keep in mind that we now have two
distinct interpolants: the $K_x$ used in real space to
define the function $F(x)$ from the samples $a_j$, and the $K_u$ used in
Fourier domain to approximate the sinc interpolation of $\tilde f(u)$
between the sample $\tilde a_j$ at $u_j=j/N$.  

Since all of our operations are linear, we can fully understand the
accumulated errors by considering the case when the input $a_j$ are
zero except for a single element $a_n=1$.  The DFT produces $\tilde
a_j = \exp(-2\pi i j n/N),$ a sine wave advancing by an amount $\nu_0
= n/N$ cycles per sample, which will be perfectly interpolated to the
expected $\tilde f(u)=\exp(-2\pi i u n)$ at any $|u|<0.5$ if we use
the sinc kernel.   

With an imperfect $K_u$, however, we can use \eqq{efunc} to infer the
error in the interpolated sine wave, putting $\nu_0=n/N$ and $t=uN$: 
\begin{eqnarray}
\tilde f(u)  & \rightarrow  & \left[1-E_0(n/N)\right] \tilde f(u) + \sum_{j\ne 0}
\tilde K_u(j+n/N) e^{2\pi i jNu} \tilde f(u) \\
E_0(\nu) & \equiv & \sum_{j\ne 0} \tilde K_u(j+\nu) = \sum_{j>0}
\left[\tilde K_u(j+\nu) + \tilde K_u(j-\nu)\right].
\end{eqnarray}
Upon transforming back to real space, the first term will yield a
scaled version of the input function, while the sum becomes a series
of ghost images at distance $jN$ from the original point:
\begin{equation}
f(x)  \rightarrow   \left[1-E_0(n/N)\right] \delta(x-n) + \sum_{j\ne 0}
\tilde K_u(j+n/N) \delta(x-n-jN).
\end{equation}
Now we may generalize to an arbitrary configuration of input samples
$a_j$.  The $\hat f(x)$ that is obtained by executing an inverse
transform after interpolation of the $\tilde a_j$ will be
\begin{equation}
\hat f(x) =  \left[1-E_0(x/N)\right] f(x) + \sum_{j\ne 0} \tilde
K_u(j+x/N) f(x-jN).
\end{equation}
These alterations to $f(x)$ will be essentially preserved when
convolved with $K_x$ to produce the function $F(x)$ and its approximation
$\hat F(x)$, namely:
\begin{itemize}
\item The approximated version will be multiplied by the function
  $1-E_0(x/N)$.
\item A series of ghosts images appear, displaced by $jN$ from $F(x)$,
  and each multiplied by the function $\tilde K_u(j+x/N)$.  
\end{itemize}
We note further that $|x/N|<0.5$, and we are choosing interpolants
intended to approximate the band-limited behavior $\tilde K(\nu)=0$
for $|\nu|>0.5$, so it is generally true that $|\tilde K_u(j\pm x/N)|
\ll |\tilde K_u(1\pm x/N)|$ for $j>1$.  In conditions considered here,
the first ghost image dominates.

\begin{figure}
\plottwo{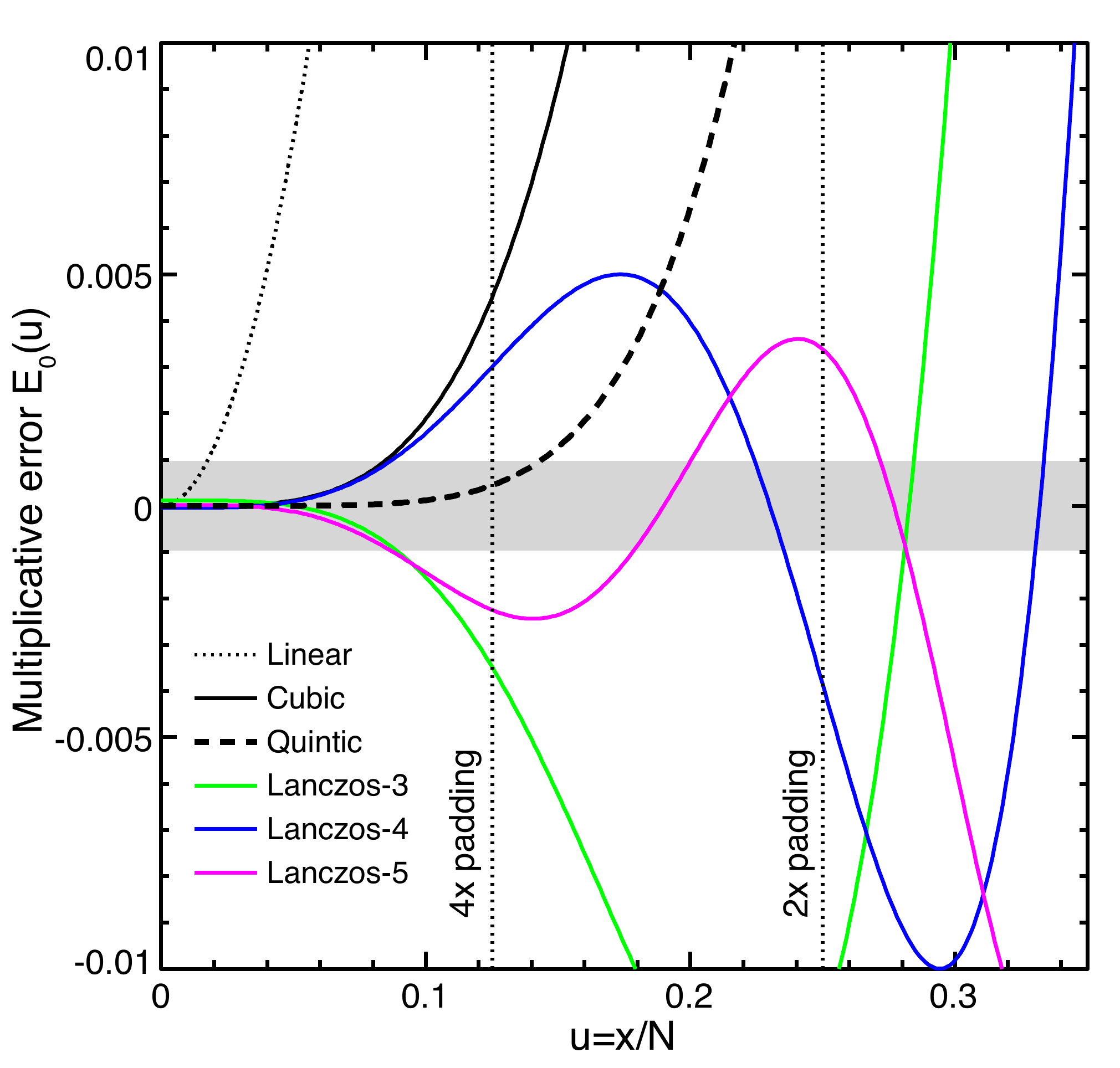}{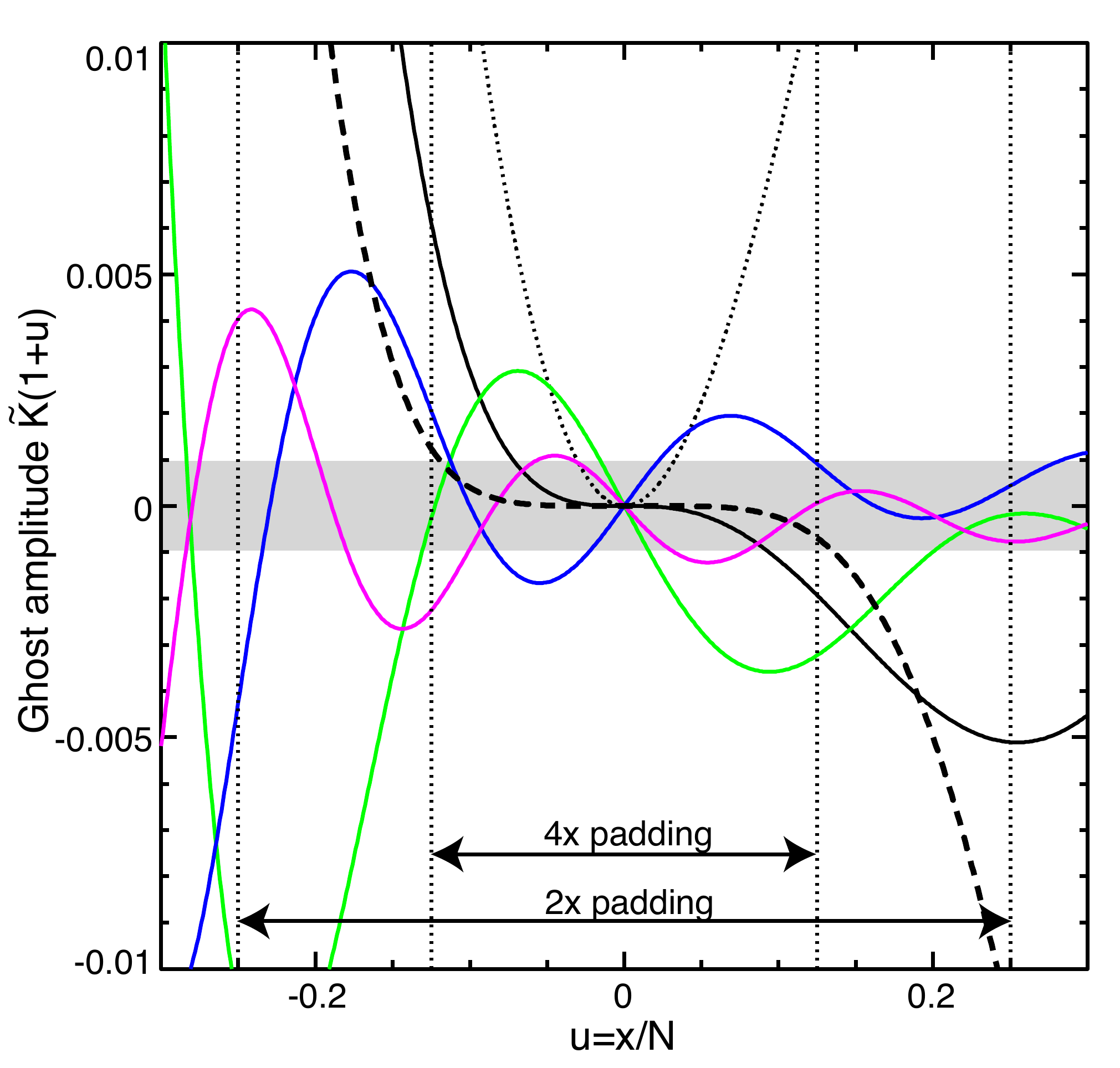}
\caption[]{Error functions for several common interpolants.  The left
  panel plots the multiplicative error $E_0(x/N)$ induced by
  approximation of the Fourier-domain sinc interpolation with the
  selected interpolant.  The right panel plots the function $\tilde
  K(1+x/N)$ that gives the amplitude of the first, dominant, ghost
  image that is produced by using the chosen $u$-space interpolant.
  The vertical dotted lines show the maximum $x/N$ that will be
  present when the input data are zero-padded by a factor 2 or 4
  before initial DFT.  The horizontal grey band shows, for reference,
  errors of $\pm1$ part per thousand.  The polynomial filters (in
  black) obtain lower errors than the Lanczos filters (in colors) for
  given kernel size if $|u|<0.125$ is enforced by 4-fold zero padding.
}
\label{E0}
\end{figure}

To quantify the first effect, the left panel of
Figure~\ref{E0} plots the function $E_0(u)$ that describes the
multiplicative errors in the central image of the reconstructed $\hat
F(x)$, for the interpolants cataloged in the Appendix.
Table~\ref{itable} lists the interpolants, their kernel sizes, and
their error levels at chosen maximum values of $x/N$.
Clearly any of these interpolants will accrue errors $\gg1\%$ for
$|x/N|>0.3$.  We can hold $|x/N|<0.25$ by zero-padding the initial
array by a factor 2 before the initial DFT.  The Lanczos interpolants
are designed to minimize $\tilde K(\nu)$ for $|\nu|>0.5$ and hence
perform well in reducing the interpolation errors determined by
$\tilde K(\nu)$ for $0.75<|x/N|<1.25$.
The Lanczos filters of
order $\ge4$ attain $|E_0(x/N)|<0.005$ with 2-fold padding.

Equivalent precision is available from the smaller cubic
interpolant kernel if we precompute the DFT with 4-fold zero-padding
to keep $|x/N|<\frac{1}{8}$.  In this range we note that the cubic
interpolant performs just as well as the Lanczos interpolants with
larger (hence slower) kernels.  This is attributable to the cubic
interpolant being designed to be exact for polynomials of order $\le
2$, which is equivalent to the requirement that $(d/d\nu)^m \tilde
K(\nu)=0$ for integral $\nu$ and $m\le 2$.  Thus the cubic interpolant
has $\tilde K(\nu)$ smaller than a Lanczos interpolant of equal size
if we stay sufficiently close to integer values of $\nu$.

Inspired by the success of the standard cubic interpolant, we seek an
interpolant that is exact for polynomial functions beyond quadratic order.  This
will require a 6-point interpolant, which must in turn be a
piecewise-quintic polynomial function.  This quintic filter, described
in the Appendix, also has continuous second derivatives.  The
quintic interpolant indeed satisfies our expectation of performing
extremely well if we confine the data to $\nu=x/N<\frac{1}{8}$ by a 4-fold
zero padding of the data before the DFT.  The multiplicative error
$E_0(x/N)$ is $<5\times10^{-4}$ in this case.

As an illustration of the second effect, the right panel of
Figure~\ref{E0} plots $\tilde K(1+u)$ for the interpolants, which
determines the relative brightness of the first ghost image.  We find
essentially the same criteria on padding and choice of interpolant as
we did from the threshold we imposed on $E_0$. This is assured, as  $E_0(\nu)\approx\tilde K(1-u)+\tilde
K(1+u)$, given that all filters beyond the
linear one have  $|\tilde K(2+u)|\le0.001$ for $|u|<0.25$.
With 4-fold zero-padding, the polynomial filters are again the most
efficient at reducing the amplitude of the ghost images.
The 4-element cubic filter limits the ghost amplitude to 0.006 of the
original $a_j$, and the 6-element quintic filter reduces the ghost
amplitude to 0.0012.

Limiting both the amplitude of the ghost images and the multiplicative
error on the central image of $\hat F(x)$ to be $<0.001$ is hence
achieved by 4-fold zero padding of the input array, with a DFT
followed by interpolation in $u$-space with the 6-point quintic
interpolant.  Similar performance is possible with the more compact
4-point cubic interpolant if the initial DFT has 6-fold zero padding.

The $O(N^4)$ computational bottleneck at step \ref{convolvestep} of
our recipe can hence be replaced by $k^2N^2$ operations in convolution
with an interpolant with $k \times k$ footprint, where $k=6 (4)$ for
the quintic (cubic) interpolant.  The oversampling by factor $s=4 (6)$ increases the time
required for the input DFT (step \ref{dftstep}) to $O(s^2N^2 \log sN)$,
leaving the DFT as the likely slowest step of the process.  However
the whole operation now scales, within constant/logarithmic factors, as
the number of pixels in the input galaxy images.  We note that if we are
rendering many versions of a given input galaxy, it may be efficient
to pre-compute and cache the oversampled input DFT.  If storage space
for this cache is not an issue, then the cubic interpolant with 6-fold
DFT padding may be faster than quintic interpolant with 4-fold DFT.

\begin{figure}
\plotone{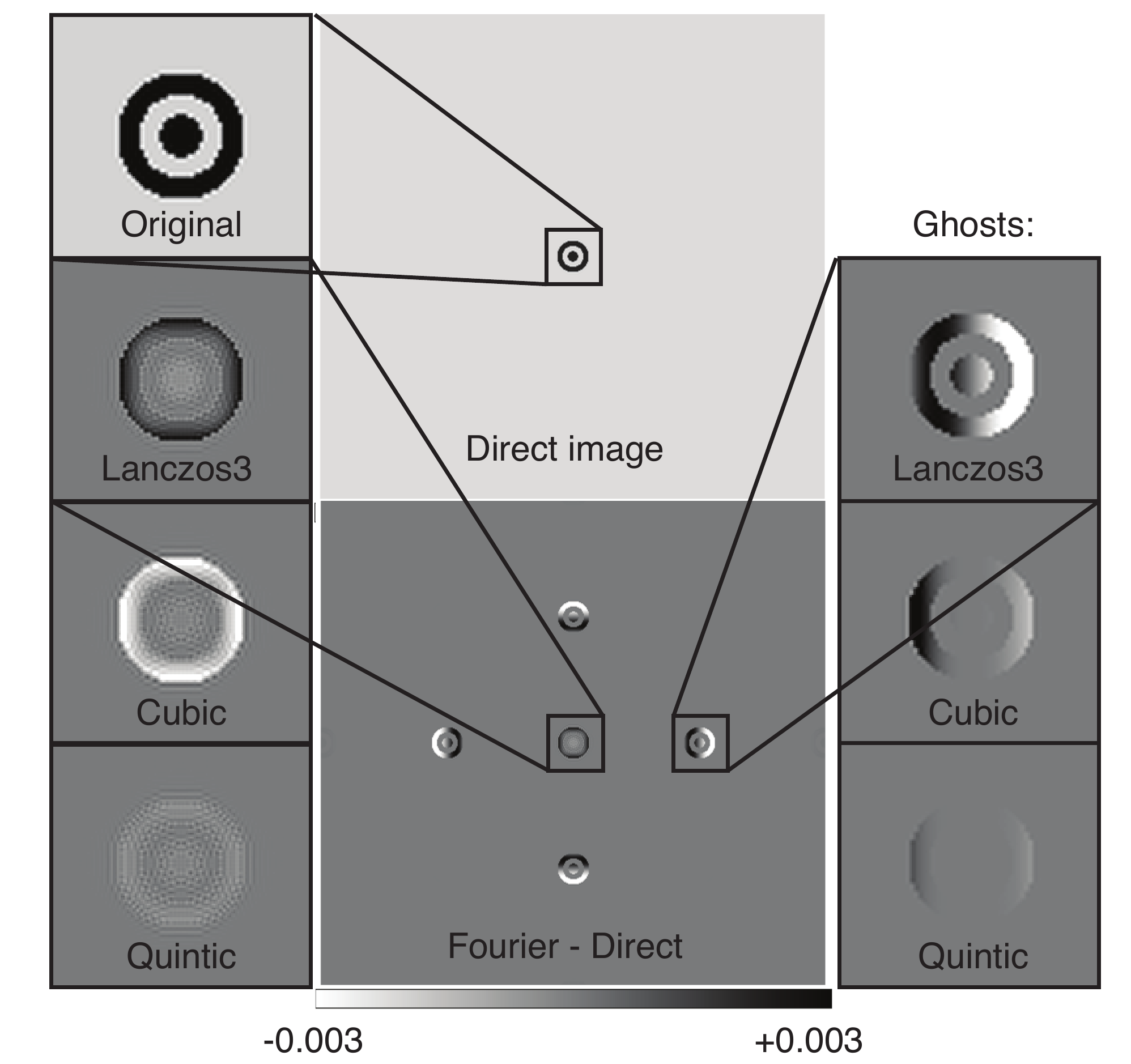}
\caption[]{Errors induced by interpolation approximations in Fourier
  domain reconstruction of a 2d image.  The original image is a
  $32\times32$ pixel bullseye pattern with unit amplitude. The upper
  middle image, with close-up in upper left, is a $2\times$
  oversampled rendition using a 3rd-order Lanczos function as the
  $x$-domain interpolant $K_x$.  
  The original image is then zero-padded to $128\times128$ pixels,
  DFT'ed to Fourier domain, interpolated to a new grid in Fourier
  domain with $K_u$ as a 3rd-order Lanczos filter, and transformed back to real
  space.  The bottom middle panel is the difference between this
  reconstruction and the direct-interpolation image, showing the
  scaling error on the central image and the 4 dominant ghost images.  The greyscale
  on the error images extends to $\pm0.003\times$ the input image's
  brightness.   At left is the error in the central image; at right is
  a closeup of the right-hand ghost image.  Changing the
  Fourier-domain interpolant from the $6\times6$ Lanczos3 kernel, to a
  $4\times4$ cubic interpolant, to a $6\times6$ quintic interpolant
  leads to progressively smaller interpolation errors.
}
\label{ghosts}
\end{figure}

Figure~\ref{ghosts} shows the errors induced in reconstruction of a
2d bullseye image using three different $u$-space interpolation
schemes.  

\begin{deluxetable}{lccccc}
\tablewidth{0pt}
\tablecolumns{8}
\tablecaption{Properties of Interpolants\label{itable}}
\tablehead{
\colhead{Name} & 
\colhead{$N_{\rm points}$} & 
\colhead{$u_{\rm max}:$} & 
\multicolumn{3}{c}{Max reconstruction error for oversampling:} \\
 & & & $2\times$ & $4\times$ & $6\times$  \\
 & & & $x/N<1/4$ & $x/N<1/8$ & $x/N<1/12$ 
}
\startdata
 Nearest & 1 &  317.5 & (large) & (large) & (large) \\[3pt]
Linear & 2 & 9.6 &  0.18 & 0.049 & 0.022 \\
Cubic & 4 & 2.74 & 0.061 & 0.0061 & 0.0016 \\
Quintic & 6 & 3.62 & 0.037 & 0.0012 & 0.00015 \\[3pt]
Lanczos $n=3$ & 6 & 1.49 & 0.014 & 0.0035 & 0.0035 \\
Lanczos $n=4$ & 8 & 1.35 & 0.005 & 0.0030 & 0.0019 \\
Lanczos $n=5$ & 10 & 1.08 & 0.004 & 0.0022 & 0.0012 \\[3pt]
Sinc & $\infty$ & 0.5 & 0 & 0 & 0 \\
\enddata
\tablecomments{Interpolants are defined in the Appendix.  $N_{\rm
    points}$ is the number of grid points per dimension summed during
  the interpolation; $u_{\rm max}$ is the largest $|u|$ for which
  $|\tilde K(u)|>0.001$, giving some estimate of the power beyond the
  Nyquist frequency $u=0.5$ induced by the interpolant. The maximum of
  $|E_0(u)|$ and $|\tilde K(1\pm u)|$ attained for $u<x/N$ is listed
  for three values of $x/N$, corresponding to $2\times$, $4\times$,
  and $6\times$ zero-padding of the data, respectively.}
\end{deluxetable}

\section{Effect of wrapping}
The above section considers the transform from $u$ space back to $x$
space to produce an output image $\hat F(x)$ to be continuous, but a
discrete transform is necessary in 
practice.  Consider this DFT to produce $M$ samples of the $x$-space
image at spacing $\Delta x$.  The DFT will sample the $u$ space at
intervals $\Delta u =  (M \Delta x)^{-1}$ to yield an output
image that has been wrapped with period $P=M \Delta x$, such that
the output point $\hat F_j = \sum_k \hat F(j\Delta x + kP)$. 
The input image $F(x)$ is confined to the extent $\pm N/2$ of the original input
array $a_j$ plus the radius of the non-zero region for the interpolant
$K_x$.   Hence as long as the output DFT has extent $P>N$ that would
contain the input image, the wrapped regions are nominally zero and
the DFT yields an exact sampled representation of the output image.

However we must recall that, in the case where we have used an
approximate interpolant $K_u$, ghost images are present in the output image
at the
locations $\pm N_x=\pm sN$, where $N_x$ is the size of the $x\rightarrow u$ input DFT, and
$s$ is the zero-padding factor applied to the input data array with
$N$ samples.  If
$P=N_x/n$ for some integer $n$, then the principal ghosts will be folded directly
atop the primary image.  This might seem damaging, but in fact means
that the reconstructed $\hat F_j$ will be {\em better}, in fact {\em perfect},
because the summed ghost images will exactly cancel the
multiplicative error $E_0(x/N_x)$ that affects the primary image.
Another way to see this is that if $P=N_x/n$, then the $u\rightarrow x$ DFT is
using exactly the $u$ values produced by the $x\rightarrow u$ DFT, and no
interpolation is being done at all.

\begin{figure}
\plotone{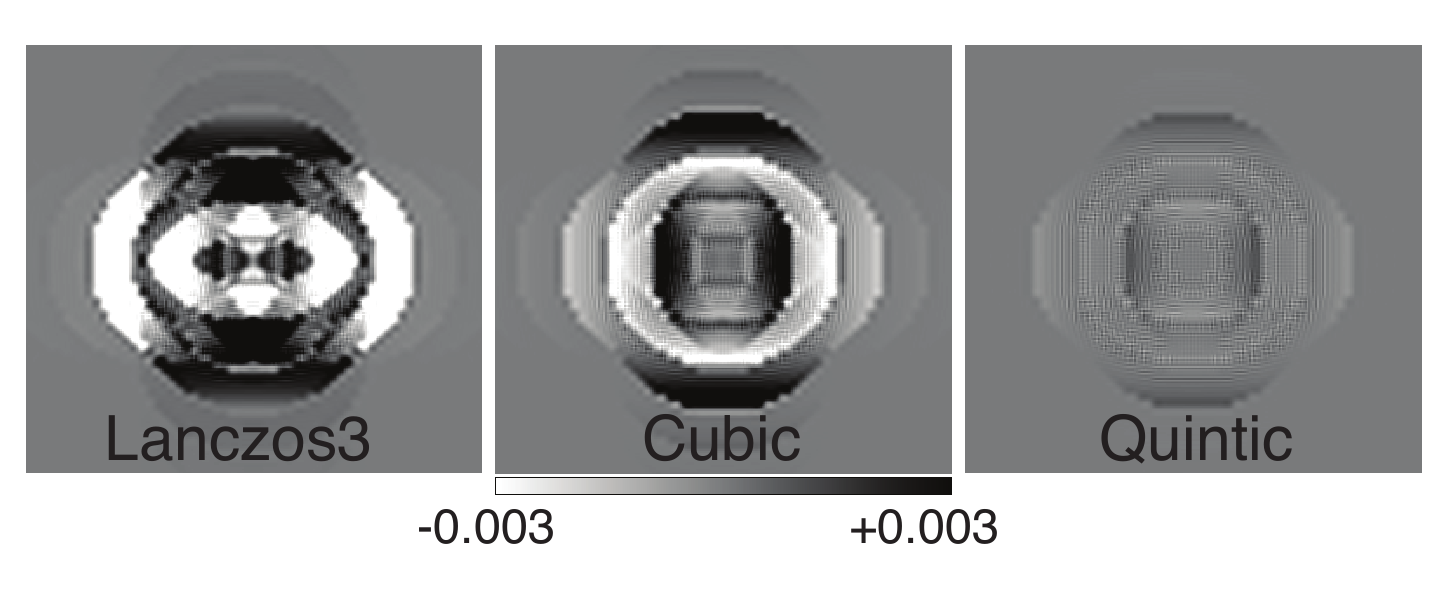}
\caption[]{
Errors induced by Fourier-domain interpolation in the case that we are
simulating a shear of amplitude 0.1 on the bullseye image.  The output
DFT is done with period $P=128$ (equal to the zero-padded input $N_x$)
such that the ghost images are wrapped atop the central bullseye, and these images show the resultant errors
of the central image in closeup.  The residuals are dominated by 4
copies of the ghost image, slightly shifted in the $\pm x$ and $\pm y$
directions because of the shear applied before the $u \rightarrow x$
DFT.  The greyscale spans $\pm0.003$ of the input bullseye
brightness.  The quadrupole patterns from the folded ghosts can
confuse weak-lensing shear measurements in this case of unfortunate
choice of DFT period, but the quintic filter is a great improvement on
the Lanczos and cubic interpolants.
}
\label{sheared}
\end{figure}

Wrapping of the ghost images can be a major problem, though, in the
following important application: suppose our goal is to take $F(x)$,
dilate it to $G(x)=F[x/(1+\epsilon)]$, then convolve with some PSF
function.  This is, for example, exactly what one wants to do to
simulate a sky that has been sheared by weak gravitational
lensing---the image must be dilated by $\epsilon=\gamma$ in one
direction and contracted with $\epsilon=-\gamma$ along a perpendicular
axis.  If we wish to execute this dilation in the Fourier domain,
setting $\tilde G(u) = \tilde F[(1+\epsilon)u]$, then the ghosts will appear at locations $x \approx
\pm(1-\epsilon)N_x$.  If we do the $u\rightarrow x$ DFT with $P=N_x$, as
might be common, then the ghosts are wrapped to positions $x=\pm
\epsilon N_x=\pm \epsilon s N$, just slightly displaced.
Figure~\ref{sheared} illustrates the results.

These displaced ghosts can generate spurious broadening of the
reconstructed image $\hat G_j$.  This is a problem when simulating sheared
2d sky images, because the interpolation errors induce a quadrupole
moment atop that induced by the shearing of the image.  We
take our ``bullseye'' galaxy and calculate the
ellipticity from unweighted quadrupole moments of the image:
\begin{eqnarray}
M_{xx} & \equiv & \sum_{jk} \hat G_{jk} x_j x_k \\
M_{yy} & \equiv & \sum_{jk} \hat G_{jk} y_j y_k \\
e & \equiv & \frac{M_{xx}-M_{yy}}{M_{xx}+M_{yy}}.
\end{eqnarray}
The original bullseye pattern has a diameter of 32 pixels, and is
$4\times$ zero-padded to 128 pixels before the $x\rightarrow u$ DFT.  We construct a sheared
image of this galaxy by interpolating in $u$ space and executing the
$u\rightarrow x$ DFT
on a $512\times512$ grid with $\Delta x=0.25$.  
The quadrupole moment of this reconstruction is compared to that of an
image constructed by direct interpolation of the $x$-space image with
a 3rd-order Lanczos kernel.  If the applied shear produces an
ellipticity $e$ on the $x$-domain interpolation of the sheared image,
we find that the reconstruction from cubic $u$-space interpolation
produces an ellipticity that is systematically biased by about
$0.04e$.  The quintic filter is better, with biases of
$\approx0.004e$. If we increase the input zero-padding from $4\times$
to $6\times$, we find the quintic filter reduces spurious
ellipticities to $<0.001e$, sufficient for highest-precision simulation
of applied shear.

Note that the bullseye image is a worst case in the respect that it
has full amplitude out to the edge of the initial domain.  In practice
we can expect postage-stamp images of galaxies to have near-zero flux
at their edges.  Since the ghosts of the quintic interpolation rise
very rapidly at the edges, a typical galaxy or PSF cutout that has
very little flux at the borders will have lower spurious contribution
from wrapping of the ghost images.

\section{Summary}
We seek a recipe for performing high-precision Fourier-domain image simulation and
analysis of galaxies and PSFs given as pixelized data, with values
between the pixel samples specified by some interpolant $K_x$.  

Our first question was: what is the exact expression
for the Fourier transform $\tilde F(u)$ of such an interpolated, sampled image?
\eqq{exact} gives the answer: first calculate the $\tilde a_k$ from a
DFT of the input samples; then interpolate between the $u_k=k/N$ using
a wrapped sinc function; then multiply by the transform $\tilde
K_x(u)$ of the $x$-domain interpolant.  The maximum frequency $u_{\rm
  max}$ with non-zero $\tilde F(u)$ hence depends on the choice of
interpolant: $K_x={\rm sinc}(x)$ provides strictly band-limited
$u_{\rm max}=0.5$ but produces an $F(x)$ that extends to very large
distance beyond the original samples.  The Lanczos interpolants are a
good choice to define $F(x)$ that is not much larger than the original
samples, while not extending $\tilde F(u)$ far beyond the original
Nyquist frequency.

The sinc interpolation of the $\tilde a_k$ is often too slow
for practical implementation at $O(N^4)$, so how could we efficiently obtain the
values $\tilde F(u)$ at arbitrary $u$ that are needed to implement
simulation of sheared and convolved renditions of $F(x)$?   
Using a $u$-space interpolant $K_u$ that with a more compact kernel
produces two errors in the simulated images: the first is a
multiplicative error $E_0(x/N)$, where $N$ is the size of the
$x\rightarrow u$ DFT and $E_0$ is a function characteristic of the
interpolant $K_u$.  The second error is the appearance of a pair of
ghost images located $\pm N$ units from the original image, each ghost
multiplied by the function $\tilde K_u(1\pm x/N)$, which is nearly the
same size as $E_0(x/N)$.

Figures~\ref{E0} and Table~\ref{itable} show the size of these errors
for common interpolants.  We find that part-per-thousand accuracy on
the resampled image and its ellipticity will be realized by the following recipe:
\begin{enumerate}
\item Zero-pad the input data by a factor 4 before performing the $x
  \rightarrow u$ DFT.
\item Use the $6\times6$ pixel piecewise-quintic-polynomial interpolant in $u$ space.
\end{enumerate}
We therefore recommend use of the quintic filter for $u$-space
interpolation after zero-padding and DFT.  The cubic filter can be
used to gain a factor 2--3 in speed at expense of $\approx 5\times$
larger simulation errors, and these larger errors can be eliminated by
6-fold zero-padding of the input image.

There is one important caveat to this recipe: if the simulated
$u$-domain image is transformed back to $x$ domain using a DFT with
period $P$ that nearly evenly divides $N$, then the ghosts will be folded
atop the primary image.  A simulated shear or magnification will cause
the ghosts to move relative to the primary image, which produces
spurious broadening of the image, which can change the quadrupole
moments of the image in a way that biases shear measurements.  In a
simple pessimistic trial case, we find that the standard
$4\times$padding$+$quintic recipe induces 0.4\% errors in a simple
measure of applied shear, somewhat too big for testing state-of-the-art
shear measurement techniques.  We
find it possible to
reduce the spurious shear to $<0.001$ of the applied shear by
combining the quintic $K_u$ with $6\times$ zero-padding of the initial
DFT, which should be sufficiently accurate for foreseeable
cosmic-shear simulations.  In fact this would be overkill for most
galaxy images that one is likely to be simulating, since they are
already nearly zero at their edges and hence not in need of $6\times$
padding.

The methods described herein have been implemented as the core of the
Fourier-domain image rendering code for the {\it GalSim} public-domain
sky simulation code\footnote{ {\tt
    https://github.com/GalSim-developers}} \citep{galsim}.

\acknowledgments
This work was supported by Department of Energy grant DE-SC0007901,
National Science Foundation grant AST-0908027 and NASA
grant NNX11AI25G. DG was supported by SFB-Transregio 33 `The Dark Universe' 
by the Deutsche Forschungsgemeinschaft (DFG) and the DFG cluster of excellence
`Origin and Structure of the Universe'.
We thank Rachel Mandelbaum, Barney Rowe, and
R. Michael Jarvis for their comments and efforts in producing usable
public code based on these derivations.

\appendix
\section{Function and interpolant definitions}
We adopt the following conventions for functions and Fourier
transforms:
\begin{eqnarray}
\quad \boxcar(x) & \equiv & \left\{ \begin{array}{cc}
1 & |x|<0.5 \\
0.5 & |x|=0.5 \\
0 & |x| > 0.5
\end{array} \right. \\
\shah(x) & \equiv & \sum_{j=-\infty}^\infty \delta(x-j) \\
\sinc(x) & \equiv & \frac{\sin \pi x}{\pi x} \\
{\rm Si(x)} & \equiv & \int_0^x dt\frac{\sin t}{t} \\
\tilde f(u) & = & \int_{-\infty}^{\infty} dx\, f(x) e^{-2\pi i ux} \\
f(x) & = & \int_{-\infty}^{\infty} du\, \tilde f(u) e^{2\pi i ux} 
\end{eqnarray}
The interpolants considered in this paper are defined as follows:
\begin{itemize}
\item {\bf Nearest-neighbor:} The real-space kernel is maximally
  compact, $K(x)=\Pi(x)$, but the Fourier domain $\tilde K(u) =
  \sinc(u)$ extends to infinity as $\sim(1/u)$.  
\item {\bf Linear:} Common interpolant with 2-point footprint in real
  space, and improved but still very broad Fourier behavior $\sim(1/u)^2$:
\begin{eqnarray}
K(x) & = & \left\{ \begin{array}{cc}
1-|x| & |x|\le1 \\
0 & |x| \ge 1
\end{array} \right. \\
\tilde K(u) & = & \sinc^2(u).
\end{eqnarray}
\item {\bf Cubic: } Piecewise-cubic polynomial interpolant with
  continuous first derivatives 
  designed to interpolate quadratic polynomials perfectly.  This implies $\tilde K^\prime(j)=\tilde
  K^{\prime\prime}(j)=0$ for integers $j\ne 0$.
The Fourier domain expression is analytic but too 
  complex to merit detailing.
\begin{equation}
K(x) = \left\{ \begin{array}{cc}
\frac{3}{2}|x^3| - \frac{5}{2} x^2 + 1  & |x|\le1 \\
-\frac{1}{2}|x^3| + \frac{5}{2} x^2 - 4|x| + 2  & 1 \le |x| \ge 2 \\
0 & |x|\ge 2
\end{array} \right. \\
\end{equation}
\item {\bf Quintic:} A six-point interpolant that provides exact
  interpolation of fourth-order polynomial functions,
  and hence has $\tilde K(j\pm\nu)\sim
  \nu^5$ for $j\ne 0$, can be produced from a piecewise-quintic
  polynomial kernel.  This kernel also has continuous second
  derivatives:
\begin{equation}
K(x) = \left\{ \begin{array}{cc}
1 + \frac{x^3}{12}\left( -95 + 138 x - 55x^2\right) & |x|\le1 \\
\frac{(x-1)(x-2)}{24} \left(-138+348x-249x^2+55x^3\right)  & 1 \le |x| \le 2 \\
\frac{(x-2)(x-3)^2}{24} \left(-54+50x-11x^2\right)  & 2 \le |x| \le 3 \\
0 & |x|\ge 3
\end{array} \right. \\
\end{equation}

\item {\bf Lanczos:} at order $m$ truncates the sinc filter after
its  $m$th null:
\begin{eqnarray} 
K(x) & \equiv & \sinc(x) \sinc(x/m) \Pi(x/2m) \\
\tilde K(u) & \equiv & 2m^2 \Pi(u) \ast \Pi(u/m) \ast \sinc (2mu) \\
 & = & \pi (2mu - m - 1) {\rm Si}(2mu - m - 1) - \pi (2mu - m + 1)
 {\rm Si}(2mu - m + 1)  \nonumber \\
 & & - \pi (2mu+ m - 1) {\rm Si}(2mu + m - 1) + \pi (2mu + m + 1)
 {\rm Si}(2mu + m + 1)  \nonumber
\end{eqnarray}
\citet{AbSteg} present approximations to the
{\rm Si} function which are very useful for calculating $\tilde K$ for
the Lanczos interpolants or for estimating leading-order behavior of
the interpolation error quantities described in this text.  See their
(5.2.6), (5.2.34), and (5.2.38).

As noted in \S\ref{flat}, the Lanczos filters do not conserve background
flux, but in practice can be modified to do so. The plots and table
assume that we are using background-conserving versions of the Lanczos filters.

\item {\bf Sinc:} conjugate of the nearest-neighbor filter, with
  $K(x)=\sinc(x)$ and $\tilde K(u) = \Pi(u)$.
\end{itemize}

\acknowledgements

\end{document}